%% file: main.tex
\documentclass[sigconf]{acmart}

\usepackage{url}
\usepackage{graphicx}
\usepackage{amsmath}
\usepackage{amsthm}
\usepackage{booktabs}
\usepackage{algorithm}
\usepackage{algorithmic}
\usepackage{color}
\usepackage{xcolor}
\usepackage{xspace}
\usepackage[misc]{ifsym}

\usepackage[ruled, vlined, nofillcomment, linesnumbered, algo2e]{algorithm2e}

\usepackage{booktabs}
\usepackage{multirow}
\usepackage{balance}
\usepackage{enumitem}
\usepackage{stfloats}
\usepackage{diagbox}
\definecolor{result_color}{RGB}{250,250,210}
\usepackage{bm}

\newcommand{\eg}{\emph{e.g.,}\xspace}

\newcommand{\ie}{\emph{i.e.,}\xspace}
\newcommand{\ignore}[1]{}
\newcommand{\paratitle}[1]
{\vspace{1.5ex}\noindent\textbf{#1}}
\AtBeginDocument{%
  }

\copyrightyear{2024}
\acmYear{2024}
\setcopyright{acmlicensed}
\acmConference[KDD '24] {Proceedings of the 30th ACM SIGKDD Conference on Knowledge Discovery and Data Mining }{August 25--29, 2024}{Barcelona, Spain.}
\acmBooktitle{Proceedings of the 30th ACM SIGKDD Conference on Knowledge Discovery and Data Mining (KDD '24), August 25--29, 2024, Barcelona, Spain}
\acmISBN{979-8-4007-0490-1/24/08}
\acmDOI{10.1145/3637528.3671734}
\begin{document}

%%
%% The "title" command has an optional parameter,
%% allowing the author to define a "short title" to be used in page headers.
\title[Revisiting Reciprocal Recommender Systems: Metrics, Formulation, and Method]{\texorpdfstring{Revisiting Reciprocal Recommender Systems:\\ Metrics, Formulation, and Method}{Revisiting Reciprocal Recommender Systems: Metrics, Formulation, and Method}}

%%
%% The "author" command and its associated commands are used to define
%% the authors and their affiliations.
%% Of note is the shared affiliation of the first two authors, and the
%% "authornote" and "authornotemark" commands
%% used to denote shared contribution to the research.
\author{Chen Yang}
\email{flust@ruc.edu.cn}
\affiliation{
    %  \department[0]{Gaoling School of Artificial Intelligence}
    %  \department[1]{Beijing Key Laboratory of Big Data Management and Analysis Methods}
         \institution{Nanbeige Lab, BOSS Zhipin}
      \city{Beijing}
  % \postcode{100872}
    \country{China}
}
\affiliation{
    %  \department[0]{Gaoling School of Artificial Intelligence}
    %  \department[1]{Beijing Key Laboratory of Big Data Management and Analysis Methods}
         \institution{Gaoling School of Artificial Intelligence, Renmin University of China}
      \city{Beijing}
  % \postcode{100872}
    \country{China}
}

\author{Sunhao Dai}
\email{sunhaodai@ruc.edu.cn}
\affiliation{
    %  \department[0]{Gaoling School of Artificial Intelligence}
    %  \department[1]{Beijing Key Laboratory of Big Data Management and Analysis Methods}
         \institution{Gaoling School of Artificial Intelligence, Renmin University of China}
      \city{Beijing}
  % \postcode{100872}
    \country{China}
}

\author{Yupeng Hou}
\email{yphou@ucsd.edu.cn}
\affiliation{
         \institution{University of California, San Diego}
 \city{La Jolla}
      % \city{United States}
  % \postcode{92092}
    \country{United States}
}

\author{Wayne Xin Zhao
\textsuperscript{\Letter}
}
\email{batmanfly@gmail.com}
\affiliation{
%   \department[0]{Gaoling School of Artificial Intelligence}
%   \department[1]{Beijing Key Laboratory of Big Data Management and Analysis Methods}
    \institution{Gaoling School of Artificial Intelligence, Renmin University of China}
  \city{Beijing}
  % \postcode{100872}
  \country{China}
}

\author{Jun Xu
}
\email{junxu@ruc.edu.cn}
\affiliation{
    \institution{Gaoling School of Artificial Intelligence, Renmin University of China}
  \city{Beijing}
  % \postcode{100872}
  \country{China}
}

\author{Yang Song
\textsuperscript{\Letter}}
\email{songyang@kanzhun.com}
\affiliation{%
  \institution{Nanbeige Lab, BOSS Zhipin}
  \city{Beijing}
  \country{China}
}

\author{Hengshu Zhu}
\email{zhuhengshu@kanzhun.com}
\affiliation{%
  \institution{Career Science Lab, BOSS Zhipin}
  \city{Beijing}
  \country{China}
}

\thanks{\Letter\ Corresponding author.}

%%
%% By default, the full list of authors will be used in the page
%% headers. Often, this list is too long, and will overlap
%% other information printed in the page headers. This command allows
%% the author to define a more concise list
%% of authors' names for this purpose.
\renewcommand{\authors}{Chen Yang, Sunhao Dai, Yupeng Hou, Wayne Xin Zhao, Jun Xu, Yang Song, Hengshu Zhu}
\renewcommand{\shortauthors}{Chen Yang et al.}

%%
%% The abstract is a short summary of the work to be presented in the
%% article.
\begin{abstract}

Reciprocal recommender systems~(RRS), conducting bilateral recommendations between two involved parties, have gained increasing attention for enhancing matching efficiency.
However, the majority of existing methods in the literature still reuse conventional ranking metrics to separately assess the performance on each side of the recommendation process. 
These methods overlook the fact that the ranking outcomes of both sides collectively influence the effectiveness of the RRS, neglecting the necessity of a more holistic evaluation and a capable systemic solution. 

In this paper, we systemically revisit the task of reciprocal recommendation, by introducing the new metrics, formulation, and method. 
Firstly, we propose five new evaluation metrics that comprehensively and accurately assess the performance of RRS from three distinct perspectives: 
overall coverage, bilateral stability, and balanced ranking. 
These metrics provide a more holistic understanding of the system's effectiveness and enable a comprehensive evaluation.
Furthermore, we formulate the RRS from a causal perspective, formulating recommendations as bilateral interventions, which can better model the decoupled effects of potential influencing factors. By utilizing the potential outcome framework, we further develop a model-agnostic causal reciprocal recommendation method that considers the causal effects of recommendations. Additionally, we introduce a reranking strategy to maximize matching outcomes, as measured by the proposed metrics.
Extensive experiments on two real-world datasets from recruitment and dating scenarios demonstrate the effectiveness of our proposed metrics and approach.
The code and dataset are available at: \textcolor{blue}{\url{https://github.com/RUCAIBox/CRRS}}.
\end{abstract}

%%
%% The code below is generated by the tool at http://dl.acm.org/ccs.cfm.
%% Please copy and paste the code instead of the example below.
%%



%%
%% Keywords. The author(s) should pick words that accurately describe
%% the work being presented. Separate the keywords with commas.
% \keywords{Do, Not, Us, This, Code, Put, the, Correct, Terms, for,
  % Your, Paper}
%% A "teaser" image appears between the author and affiliation
%% information and the body of the document, and typically spans the
%% page.

%%
%% This command processes the author and affiliation and title
%% information and builds the first part of the formatted document.
\maketitle

\input{sec_introduction}

\input{sec_relatedwork}

\input{sec_preliminary}

\input{sec_metric}

\input{sec_model}

\input{sec_experiment}

\input{sec_conclusion}

\begin{acks}
This work was partially supported by National Natural Science Foundation of China under Grant No. 62222215, and Beijing Natural Science Foundation under Grant No. L233008 and 4222027. Xin Zhao is the corresponding author.
\end{acks}
\balance

\bibliographystyle{ACM-Reference-Format}
\bibliography{ref}

\end{document}

%% file: sec_introduction.tex
\section{Introduction}\label{sec:intro}
Reciprocal recommender systems (RRS), refer to the task that
two-sided adoption has to be achieved for the final match, 
which has been widely studied and applied to scenarios like marriage markets~\cite{li2012meet,choo2006marries,pizzato2010recon, xia2015reciprocal}, recruitment~\cite{yu2011reciprocal, yang2022modeling}, and social networks~\cite{alsaleh2011improving, he2010social}. Distinct from conventional user-item recommender systems~\cite{resnick1997recommender,lu2015recommender}, RRS typically involves two parties of users and aims to generate mutually beneficial recommendations. 
Each party within the system exhibits unique preferences, requirements, and responses to the recommendations provided.
Despite different task settings, most studies on RRS typically follow existing evaluation paradigms~\cite{zheng2023reciprocal} (especially similar evaluation metrics), except that the evaluation process is conducted for each party independently.

Despite the simplicity and popularity, we argue that such an evaluation approach is inherently problematic since the recommendation results of both parties are coupled while the single-sided evaluation process is independent.  
Take online recruitment as an example and assume that a candidate sees the recruitment ad and the recruiter interviews and accepts him. 
When a match already occurs between the candidate and recruiter incurred by a recommendation to the candidate, there is no need to make another recommendation for the recruiter, which actually becomes redundant. 
However, in such a case, independent evaluations would consider both-sided recommendations as two successful recommendations, even though it doesn't increase the total number of matches made by the system.
In the reciprocal scenario, an ideal goal should be the total number of matched pairs~\cite{su2022optimizing}.
Thus, a more comprehensive evaluation perspective taking into account the overall number of matches and the degree of recommendation redundancy is needed for accurately measuring RRS performance.

In this work, we aim to evaluate the effectiveness of RRS models from a holistic perspective. We propose to consider three additional aspects for evaluation: the coverage of the overall matching relationship, the duplication degree of the recommendations, and \textcolor{black}{the ranking performance with population imbalance of the two sides}, which we define as \emph{overall coverage}, \emph{bilateral stability}, and \emph{balanced ranking}, respectively.
The underlying purpose of the overall coverage and bilateral stability metrics is to identify and filter \emph{redundant recommendations}, \ie pairs of users that are recommended to each other on both sides. 
The balanced ranking aims to evaluate the performance of ranking lists while ensuring equal treatment of users on both sides, thereby mitigating bilateral imbalances.
% On platforms where unilateral interaction can reflect whether the two users match, ``overall coverage'' can be a good measure of the recommendation performance, which counts how many pairs of potentially matched users have been revealed by RRS (recommended from at least one side). 
% In contrast, in scenarios where two users match only when they both like each other,  ``bilateral stability'' becomes more suited to assess if users in actually matched pairs have been mutually recommended. 

\begin{figure}[t!]
	\centering
	\includegraphics[width=0.47\textwidth]{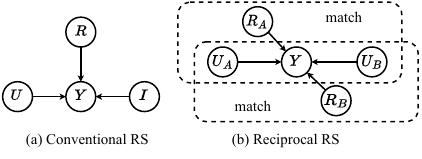}
	\caption{\textcolor{black}{Causal graph for (a) recommendation as treatment in convention RS; (b) reciprocal recommendation as bilateral treatments in RRS.
 U: user, I: item, R: recommendation, $\mathbf{R_A}$ and $\mathbf{R_B}$: recommendations made for each side,
 Y: ranking score (e.g., the probability of matching). For simplicity, we ignore all confounding factors.
 }}
	\label{fig:causalgraph}
\end{figure}

%\textcolor{blue}{
In addition to evaluating RRS from a holistic perspective, it is essential to study how recommendations in this scenario will affect the overall \textcolor{black}{effectiveness} when designing RRS models. 
Compared with conventional recommender systems, it becomes more intricate to model the effects of potential recommendations in RRSs. For example, when a matching relationship occurs with bilateral recommendations, it can be attributed to the recommendations made on either side or both sides. Without deeply understanding the underlying mechanism, it isn't easy to accurately model the two-sided matching process. 
To better understand the differences between conventional and reciprocal recommender systems, we illustrate RRS from a causal perspective~\cite{pearl2009causality} in Figure~\ref{fig:causalgraph}.
Compared to the conventional setting, the recommendations in RRS for both sides may collaboratively influence the matching outcome, offering increased likelihood for a match.
% , as shown in Figure~\ref{fig:causalgraph}~(b). 
This combined effect of bilateral recommendations can be viewed as an effect of \textcolor{black}{mixed treatments}  on the outcomes, which would affect the final results in an integrated manner. 
It motivates us to consider such a causal effect of bilateral recommendations for improving RRS. 
%

%\textcolor{blue}{
To address the above issues, we propose a model-agnostic learning approach, namely \textbf{C}ausal \textbf{R}eciprocal \textbf{R}ecommender \textbf{S}ystem (\textbf{CRRS}), formulated under the \emph{potential outcome framework}~\cite{rubin1980randomization}.
The core idea is to estimate the matching probability under different treatment \textcolor{black}{assignments}, which are termed the potential outcomes.
There are two major technical contributions in our approach.
\textcolor{black}{
Firstly, we propose an adaptable framework for fine-grained modeling of causal effects. This approach can be extended to existing typical recommendation system models, capturing the impact of different recommendation strategies on the outcomes.
}
\textcolor{black}{
Secondly, we design an effective reranking strategy for refining the ranking score for both sides. This strategy comprehensively considers the recommendation mechanism within the reciprocal scenario, aiming to enhance the overall performance of the RRS.
}

\textcolor{black}{The contributions of this paper can be summarized as follows: }

$\bullet$ We introduce a comprehensive set of metrics designed to evaluate RRS holistically, addressing aspects traditionally neglected by standard metrics in existing recommender systems.
% ~(Section~\ref{sec:metric})

$\bullet$ We take a causal perspective to model the matching effect in RRS and develop a model-agnostic framework, enabling improved prediction and user ranking by viewing reciprocal recommendations as bilateral treatments.

$\bullet$ We conduct extensive experiments to demonstrate the effectiveness of our proposed metrics and methods compared with reciprocal recommendation models on two real-world datasets.

%% file: sec_relatedwork.tex
\section{Related Work}
\label{sec:related_work}

\paratitle{Reciprocal Recommender System.}
Reciprocal Recommender Systems~(RRS) have been widely studied for two-sided markets~\cite{siting2012job, mine2013reciprocal, yu2011reciprocal, palomares2021reciprocal}, recommending users to other users rather than items.
RRS are frequently employed in domains where both sides of the market have their preferences and interactions, such as online dating~\cite{pizzato2010recon, xia2015reciprocal, tu2014online, li2012meet, pizzato2013recommending}, recruitment~\cite{mine2013reciprocal, yu2011reciprocal, su2022optimizing, yang2022modeling, chen2023bilateral, hu2023boss} and social networks~\cite{alsaleh2011improving, he2010social}.
\textcolor{black}{
Since the reciprocal recommendation task can be regarded as two separate tasks, most RRS evaluate their methods individually for each task~\cite{yang2022modeling, pizzato2010recon, neve2019latent, zheng2023reciprocal, lai2024knowledge, zheng2024bilateral}, with widely used top-$K$ recommendation metrics such as Recall, Precision, and Normalized Discounted Cumulative Gain~(NDCG). }
However, these two tasks are two sides of a coin and the final result is determined by both sides of the market.
For a better evaluation of the whole system, studies have been conducted that build upon the economic and social science literature~\cite{gale1962college, su2022optimizing, tomita2022matching, mladenov2020optimizing}, referring to this scenario as a \textit{matching market}. 
However, these methods heavily rely on the ground truth of relevance, often unavailable in real-world datasets.
In this work, we propose to evaluate RRS methods from an overall perspective, addressing overall performance across three aspects: overall coverage, bilateral stability, and balanced ranking.

\paratitle{Causal Inference in Recommendation.}
Recently, causal inference~\cite{pearl2010causal} has gained significant attention in recommender systems research~\cite{liang2016causal, wang2020causal, bonner2018causal}. The current works have explored the application of causal inference to address various challenges in recommender systems, including bias~\cite{chen2023bias, schnabel2016recommendations, zhu2023causal}, fairness~\cite{huang2022achieving, wang2023survey}, and explainability~\cite{xu2021learning, wang2022sequential}. Generally, causal inference encompasses two primary directions: causal discovery and causal estimation~\cite{yao2021survey}. Causal discovery aims to learn causal relationships from data, identifying causal graphs that reveal interdependencies among factors~\cite{hainmueller2012entropy, wang2022sequential, xu2023causal}. Causal estimation seeks to estimate the treatment effect, especially evaluating how interventions or treatments influence user outcomes in recommender systems~\cite{sharma2015estimating, xie2021causcf, sato2016modeling, gao2022causal}.
Inspired by advances in causal effect estimation, our work concentrates on applying this approach to the domain of reciprocal recommendation. In this paper, we attempt to estimate the causal effects from both sides of RRS to enhance recommendation quality.

%% file: sec_preliminary.tex
\section{Problem Definition}
\label{sec:preliminary}
In reciprocal recommender systems~(RRS), there are two involved parties (typically two sets of users) with mutual selection relationship, 
denoted as $\mathcal{A} = \{a_1, a_2, \ldots, a_n\}$ and $\mathcal{B} = \{b_1, b_2, \ldots, b_m\}$, where $n$ and $m$ denote the numbers of candidates on each side.
The two parties are associated with a set of directed user-to-user interactions represented by $\mathcal{I} = \{\langle a_i, b_j, d_{ij}, r_{ij} \rangle\}$.
Each tuple represents an interaction between a user $a_i \in \mathcal{A}$ and a user $b_j \in \mathcal{B}$, where the binary variable $d_{ij} \in \{0,1\}$ indicates the direction of the interaction (1 for $a_i \rightarrow b_j$ and 0 for $b_j \rightarrow a_i$) 
and $r_{ij} \in \{0,1\}$ represents whether a final match is achieved.

In the scenario of reciprocal recommendation, 
we perform personalized recommendations for users of both sides.
Specifically, for a given user $a_i$, the system generates a ranked list of recommendations from $\mathcal{B}$.
The other side is similar, where users from $\mathcal{A}$ are ranked when recommending to user $b_j$.
Most existing works follow conventional evaluation paradigms to assess the ranking performance of each side individually.
However, in a reciprocal scenario, the primary goal should be the overall count of matching pairs.
This work focuses on evaluating the recommendations from both sides from an overall system perspective. This includes how many actual match pairs are covered by the recommendations (coverage) and whether they are recommended to each other (stability).

%% file: sec_metric.tex
\section{Revisiting the Evaluation Metrics}
\label{sec:metric}
As discussed in Section~\ref{sec:intro},
to evaluate the performance of RRS, existing methods mainly focus on either single-sided metrics or solely the overall quantity of matches, which lack consideration of more robust two-sided measurements, such as coverage and stability. 
To ensure a more holistic and robust evaluation, we propose to consider three new evaluation aspects, namely overall coverage, bilateral stability, and balanced ranking.

\subsection{Proposed Metrics}
\paratitle{Measures of Overall Coverage.}
\label{subsec:coverage} 
Overall coverage refers to the extent to which a recommender system covers potential matching relationships from an overall perspective.  
To achieve a more accurate evaluation of the coverage in RRS, the successful matching led by the bilateral recommendation, where the two users involved are recommended to each other simultaneously, should be treated as a single successful recommendation, rather than being counted twice. 
Considering this, we propose two metrics, namely \emph{Coverage-adjusted Recall}~(CRecall) and \emph{Coverage-adjusted Precision}~(CPrecision).
The corresponding formulas are given as follows:
\begin{align}
\mathrm {CRecall} &=\frac{{\mathrm{TP}_{A}} + {\mathrm{TP}_{B}} - \mathrm{TP}_{A \cap B}}{\mathrm{M}}, \\
\mathrm {CPrecision} &=\frac{{\mathrm{TP}_{A}} + {\mathrm{TP}_{B}} - \mathrm{TP}_{A \cap B}}{(n + m)K},
\end{align}
where $\mathrm{M}$ denotes the total count of the matching pairs and $K$ denotes the length of the recommendation list.
$\mathrm{TP}_{A}=\sum_{i\in \mathcal{A}}{\mathrm{TP}_{i}}$ and $\mathrm{TP}_{B}=\sum_{j\in \mathcal{B}}{\mathrm{TP}_{j}}$ denote the sum of true positives for side $A$ and side $B$, where
$\mathrm{TP}_i$ and $\mathrm{TP}_j$ are the true positives of the single user of the two sides.
\textcolor{black}{$\mathrm{TP}_{A \cap B}$ denotes the true positives where both sides have recommendations simultaneously.} 
As we can see, CRecall and CPrecision refine the positive pairs generated by the RRS by avoiding duplicate counting of redundant successful recommendations.
% aaaaa aaaaa aaaaa aaaaa.
% With CRecall and CPrecision, we can evaluate the RRS from the perspective of overall coverage, which refers to the recommended percentage of potential matching pairs.

\paratitle{Measures of Bilateral Stability.}
Bilateral stability refers to the extent to which a recommender system simultaneously recommends a pair of users to each other.
Stability is mainly related to the number of mutual recommendations, which potentially increases the likelihood of successful matching for the involved pairs.
To effectively evaluate the bilateral stability of RRS, we propose two metrics,
namely \emph{Stability-adjusted Recall}~(SRecall) and \emph{Stability-adjusted Precision}~(SPrecision).
The formulas are given as:
\textcolor{black}{
\begin{align}
\mathrm {SRecall} &=\frac{\mathrm{TP}_{A \cap B}}{\mathrm{M}}, \\
\mathrm {SPrecision} &=\frac{\mathrm{TP}_{A \cap B}}{(n + m)K}.
\end{align}
}

SRecall and SPrecision directly show how well the system simultaneously offers recommendations for the two sides of the given pair, effectively capturing the essence of bilateral stability.

\paratitle{Measures of \textcolor{black}{Balanced Ranking}.}
\textcolor{black}{
Balanced ranking refers to the ranking performance of the recommendation list
ensuring equity across different group sizes.
}
An essential characteristic of the RRS is the bilateral imbalances and crowding~\cite{su2022optimizing}. For instance, in a recruitment scenario, there might be 1000 candidates and merely 100 jobs, resulting in a significant disparity in the number of users.
To fairly evaluate the ranking performance of each user in the RRS, we propose \emph{Reciprocal $\mathrm{NDCG}$}~(RNDCG) based on the widely-used ranking metric $\mathrm{NDCG}$~\cite{wang2013theoretical}. It offers a more comprehensive evaluation of overall ranking performance, considering weight disparities between different sides.
The formula is presented as:
\begin{align}
    \mathrm {RNDCG@K} =\frac{n \cdot \mathrm{NDCG_A} + m \cdot \mathrm{NDCG_B}}{n + m}.
\end{align}

RNDCG is a weighted sum of NDCG according to the number of users on each side.
Compared with NDCG, RNDCG enhances fairness and effectiveness in the recommendations, making the system more responsive to the actual distribution of users and their potential impact on the market dynamics.

\emph{Summary}. In the above, we have presented the metrics of \emph{overall coverage} (CRecall/CPrecision), \emph{bilateral stability} (SRecall/SPrecision), and \emph{balanced ranking} (RNDCG). 
Among these proposed metrics, CRecall/CPrecision along with SRecall/SPrecision aim to enhance the evaluation of RRS, aiming to promote more balanced and mutually beneficial RRS models. 
Furthermore, RNDCG allows us to assess the ranking performance of \textcolor{black}{unbalanced groups} while accounting for the crowding, ultimately leading to \textcolor{black}{more informed and equitable recommendations for all users. }

\subsection{Effectiveness of the Proposed Metrics}
\label{sec:metric_analysis}
\begin{figure}[t!]
	\centering
	\includegraphics[width=0.47\textwidth]{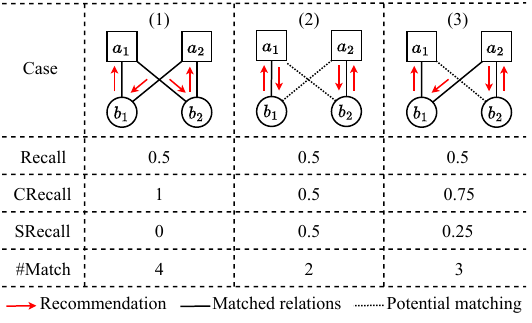}
	\caption{Three distinct recommendation cases with identical matching relationships. 
 In this scenario, we make a top-1 recommendation for four users~($a_1$, $a_2$, $b_1$, and $b_2$) and then report the average Recall, CRecall, SRecall, and the count of successful matching pairs.
 }
	\label{fig:rrsexample}
\end{figure}

To better understand the advantage of the proposed metrics,  
we construct three illustrative scenarios as depicted in Figure~\ref{fig:rrsexample}. For clarity, we consider a top-1 recommendation scenario, where each user receives just one recommendation (indicated by arrows). Here, matches cannot be realized without recommendations~(namely potential matching, indicated by black dashed lines). 

The three scenarios represent distinct recommendation strategies. 
In case 1, four separate recommendations each successfully established a matched relation. 
In case 2, two mutual recommendations resulted in two matched relations while in case 3, three matched relations were successfully established.
We calculate both traditional and proposed metrics for these cases. 
For simplicity, we take Recall, CRecall, and SRecall for comparisons, and the other metrics can be conducted similarly. 
It is evident that traditional metrics fail to capture the differences among the three cases (all having Recall=0.5), whereas our proposed metrics CRecall and SRecall can capture such a distinction~(\eg CRecall=1 in case~1 and CRecall=0.5 in case~2).
Thus, these proposed metrics can provide a more comprehensive evaluation of the RRS, surpassing the limitations of traditional metrics.

In traditional RRS, when a pair is recommended by both sides and achieves a match, it is unclear through which mechanism the match was realized. 
This uncertainty introduces challenges in understanding the effectiveness of the recommendation process, making it difficult to enhance and refine the system's performance.

%% file: sec_model.tex
\section{Proposed Method}
\label{sec:model}

\begin{figure*}[t!]
	\centering
	\includegraphics[width=1\textwidth]{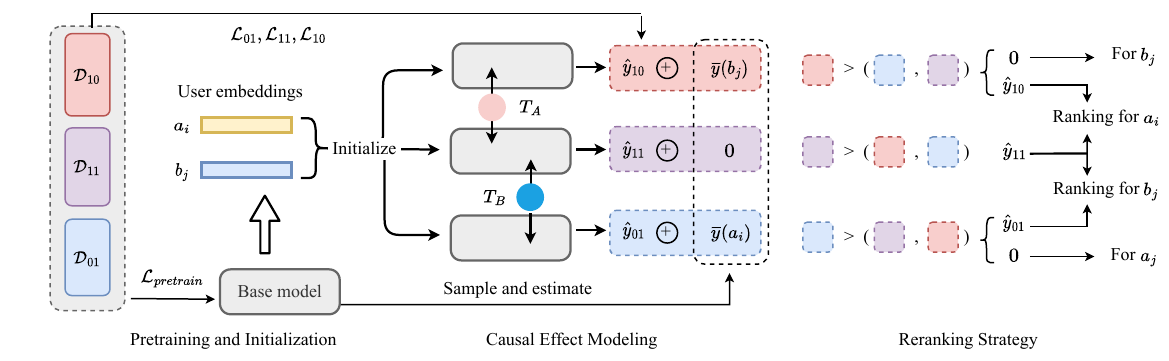}
	\caption{\textcolor{black}{An illustration of the proposed framework CRRS. 
    We represent three treatment assignments with red, purple, and blue. Each assignment signifies a distinct dataset and outcome obtained under the treatment conditions. 
 }}
	\label{fig:model}
\end{figure*}

In the above, we have introduced several new evaluation metrics for improving the evaluation of RRS. Next, we study how to develop more effective reciprocal recommendation models, so as to achieve better performance on RRS.  
Compared with conventional recommendation tasks, reciprocal recommendation becomes more intricate in the matching mechanism, due to the potential effects of two-sided recommendation. 
In order to achieve a better recommendation performance, it is key to effectively model and analyze the effects of the potential factors in deriving a matching relationship.  
For this purpose, we formulate the reciprocal recommendation scenario from a causal perspective and then introduce the proposed \textbf{C}ausal \textbf{R}eciprocal \textbf{R}ecommender \textbf{S}ystem (named \textbf{CRRS}) under the potential outcome framework~\cite{rubin1980randomization}.
The overall architecture of the proposed approach is shown in Figure~\ref{fig:model}.

%\textcolor{blue}{
\subsection{Recommendations as Bilateral Treatments}
As previously mentioned, the primary goal of an RRS is to optimize the coverage of successful matching relationships. 
This goal is highly affected by the recommendation process, either single-sided or two-sided recommendations. Given a successful matching with bilateral recommendations, it is intricate to infer which recommendation actually leads to the matching. 
To better capture the underlying effects of potential factors, we view the recommendations as treatments from a causal perspective~\cite{schnabel2016recommendations, wang2023causal}. 
In this scenario, treatments consist of two parts: recommendations for side~$A$~(denoted as $T_A$) and recommendations for side~$B$~($T_B$). 
Formally, we define \textit{bilateral treatment} as follows:
\begin{definition}[Bilateral Treatments] 
    For a given user pair $(a_i, b_j)$, $T_A$ represents whether the system recommends user $b_j$ to $a_i$, while $T_B$ represents the recommendation for the other side, \ie
\begin{align}
T_A = 1 \leftrightarrow \text{recommending}~~{b_j}~~\text{to}~~a_i, \\
T_B = 1 \leftrightarrow \text{recommending}~~{a_i}~~\text{to}~~b_j. 
\end{align}
\end{definition}
The coupling of these two distinct treatments influences the outcomes profoundly and results in four unique instances of intervention~(\ie $T_AT_B \in \{10, 11, 01, 00\}$). Based on the causal formulation, we next estimate the causal effect of each treatment combination and calculate the ranking scores for users.

\subsection{Causal Reciprocal Recommender System}
To decouple these treatments and model them separately at a finer granularity,
we leverage the \emph{potential outcome framework}~\cite{rubin1980randomization} to analyze and estimate the causal effect.
The core idea is to \textcolor{black}{estimate the causal
effect of recommendation in a collaborative way, under the assumption that similar users have similar treatment effects under recommendations~\cite{xie2021causcf}}.
By developing a collaborative causal solution, our approach is model-agnostic, ensuring adaptability and versatility.
After the causal effect is estimated, we need to determine the final ranking list presented to both sides' users.
To improve the overall \textcolor{black}{effectiveness} of the system, we further propose a reranking strategy aimed at enhancing overall performance and choosing the best recommendation strategy.

\subsubsection{Potential Outcome Framework}
In causal inference, potential outcome framework~\cite{rubin1980randomization} is a method that directly calculates the effect of certain treatments for outcomes. 
In recommendation scenarios,  \emph{outcome} is often defined as 
either implicit or explicit user feedback, such as click or rating~\cite{xu2023causal}.
%user behavior~(\eg click, purchase) or user preference~(\eg rating). 
When considering a reciprocal recommendation scenario, we define outcomes as matching relationships, as this is the most effective feedback observed.
Formally, we give the definition of \emph{potential outcome} in RRS as follows:
\begin{definition}[Potential Outcome for RRS] 
    Given the treatment assignment $T = (T_A, T_B)$, the potential outcome $Y_T$ denotes matching relationship under the specific recommendation strategy where $T = t \in \{10, 11, 01, 00\}$ for instance user pair $(a_i, b_j)$.
    The potential outcomes can be instantiated as a function $Y_T = f_T(\bm{x}_i, \bm{x}_j)$, which takes 
    % each pair's treatment assignment $t$, 
    user features $\bm{x}_i$ and $\bm{x}_j$  as input, \ie $\hat{y}_t = f_t(\bm{x}_i, \bm{x}_j)$.
    % where $t$ denotes each pair's treatment assignment.
\end{definition}

\begin{table}[t]
    \centering
    \caption{{An illustration of the expectation of matching with the bilateral treatments.
    Each unit comprises two parts: the expected outcome of the intervention and the expected outcome of other recommendations in the vacant slots.}} 
    \label{tab:pof}
    \footnotesize
    \resizebox{0.47 \textwidth}{!}{
    \begin{tabular}{c|c|c|c|c}
    \toprule
    \multicolumn{2}{c|}{\textbf{Treatment}} & {\textbf{Outcome}} & \multicolumn{2}{c}{\textbf{Matching Expectation with Vacant Slots}} \\
    % \midrule
    $T_A$ &  $T_B$ & $\hat{y}$ &{\quad \quad \quad $A$ side} \quad \quad \quad& {\quad \quad $B$ side} \quad \quad \quad\\
    \midrule
    1 & 0 & $\hat{y}_{10}$ & \quad \quad \quad 0 \quad \quad \quad & \quad \quad$\bar{y}(b_j)$ \quad \quad \quad\\
    1 & 1 & $\hat{y}_{11}$ & \quad \quad \quad 0 \quad \quad \quad& \quad \quad 0 \quad \quad \quad\\
    0 & 1 & $\hat{y}_{01}$ & \quad \quad \quad  $\bar{y}(a_i)$ \quad \quad \quad & \quad \quad 0 \quad \quad \quad\\
    0 & 0 & $\hat{y}_{00}$  & \quad \quad \quad $\bar{y}(a_i)$ \quad \quad \quad & \quad \quad $\bar{y}(b_j)$ \quad \quad \quad\\
    % \textbf{Recommend $a_i$} & $\hat{y}_{11}$   & $\hat{y}_{01} + \bar{y}(a_i) $  \\
    % \textbf{Not recommend $a_i$} & $\hat{y}_{10} + \bar{y}(b_j) $   & $\hat{y}_{00} +\bar{y}(a_i) + \bar{y}(b_j) $  \\ 
    \bottomrule
    \end{tabular}
    }
\end{table}

We present the correspondence of $\hat{y}$ to treatment $t$ in Table~\ref{tab:pof}.
Specifically, an estimation of $\hat{y}_{10} = 1$ indicates that a match is expected when the treatment combination is assigned as $T_A = 1$ and $T_B = 0$. 
Conversely, the estimation of $\hat{y}_{00}$ consistently remains at 0 because no match is expected when neither side receives a recommendation. 
To model these bilateral treatments, a major challenge is that there is only one intervention per sample in RRS.  
We cannot observe the final results of different treatments for the same pair of users at the same time, \textcolor{black}{which is a counterfactual problem~\cite{xu2023causal}.}
Therefore, it is challenging to estimate counterfactual outcomes and calculate the matching probability for reciprocal recommendations.
However, based on the intuition that similar users correspond to similar treatment effects~\cite{xie2021causcf}, we can estimate the causal effect of 
three different treatment assignments with typical RS models.

\subsubsection{Causal Effect Modeling and Ranking}
\label{sec:threemodel}
As discussed before, we reformulate the RRS scenario with the potential outcome framework and regard bilateral recommendations as treatments. 
Under the four treatment assignments mentioned above, the cases that we have to consider that are likely to be successful are those with $t \in \{10, 11, 01\}$.
As shown in Table~\ref{tab:pof}, $t= 10$ is to recommend $b_j$ to $a_i$, $t=01$ is to recommend $a_i$ to $b_j$, and $t=11$ is to recommend both users to each other.
When $t=00$, the pair cannot be matched because neither side is recommended to the other.
To support this fine-grained modeling and predict the potential outcomes under the three treatments, we construct three functions:
\begin{align}
\hat{y}_{10} = f_{10}({a}_i, {b}_j), ~~~~
\hat{y}_{11} = f_{11}({a}_i, {b}_j), ~~~~
\hat{y}_{01} = f_{01}({a}_i, {b}_j),
\end{align}
where each $f_t(\cdot)$ represents a specific recommender model backbone and each predicted value ranges from 0 to 1.
The three backbone models essentially have few restrictions and can be implemented by typical recommendation models
(\eg BPRMF~\cite{rendle2009bprmf} or LightGCN~\cite{he2020lightgcn}).   
Nonetheless, they possess unique parameters, which allow them to outperform a single model in effectively predicting outcomes for different treatments.
Our proposed method is model-agnostic and can be naturally applied to typical recommendation models.
Following this paradigm, the expectations of matching under different recommendation strategies can be estimated.

After the three outcomes have been predicted, we need to rank the candidates for each user on both sides.
For a given user pair $(a_i, b_j)$, the scores for ranking can be simply calculated using the following formula:
\begin{align}
s_{a_i} = \hat{y}_{10} + \hat{y}_{11}, ~~~
s_{b_j} = \hat{y}_{01} + \hat{y}_{11},
\end{align}
where $s_{a_i}$ and $s_{b_j}$ are the scores of recommendations for $a_i$ and $b_j$, respectively. 
Each score is the aggregate of two scores calculated under the treatment for which the recommendation is provided, either as a direct sum or alternatively, as a weighted sum.

\subsubsection{Reranking Strategy with Vacant Slots}
Although the aforementioned approach facilitates ranking, it still falls short of fully addressing the recommendation mechanism of this scenario.

In practice, if one side of a pair (\eg $a_i$ and $b_j$) lacks recommendations (\eg $T_A = 0$), there will be an empty spot for the recommendation and we term this a \emph{vacant slot}.
Another user (say $b_v$) will be recommended to $a_i$ at this vacant slot.
Such a recommendation of $b_v$ introduces a new possibility of matching denoted as $\bar{y}(a_i)$, which is the matching expectation of the pair~($a_i$, $b_v$).
Consequently, the global expectation of matching becomes $\hat{y}_{01} + \bar{y}(a_i)$. 
The results under other treatments are shown in Table~\ref{tab:pof}.
With the potential outcome predictions and matching expectations brought by the vacant slots, we need to choose and refine the scores for ranking users for the two sides.
The formula for ranking scores for the two sides is set as follows:
\begin{align}
\label{eq:reranking}
s_{a_i} &= \text{IsMax}(\hat{y}_{11}) \cdot  \hat{y}_{11} + \text{IsMax}(\hat{y}_{10} + \bar{y}(b_j)) \cdot \hat{y}_{10}, \\
s_{b_j} &= \text{IsMax}(\hat{y}_{11}) \cdot  \hat{y}_{11} + \text{IsMax}(\hat{y}_{01} + \bar{y}(a_i)) \cdot \hat{y}_{01},
\end{align} 
where $\text{IsMax}(\cdot)$ is a boolean value for the set $\mathcal{S} = \{\hat{y}_{11}, \hat{y}_{01} + \bar{y}(a_i), \hat{y}_{10} + \bar{y}(b_j), \bar{y}(a_i) + \bar{y}(b_j)\}$, denoting the best recommendation strategy for a given pair:
\begin{align}
\text{IsMax}(y) = \begin{cases} 
1 & \text{if } y = \max(\mathcal{S}), \\
0 & \text{otherwise}.
\end{cases}
\end{align}

The aim of $\text{IsMax}$ is to ensure that the final recommendation method is targeted at maximizing overall matching performance.
To estimate $\bar{y}(a_i)$ and $\bar{y}(b_j)$, we sample a certain number of users from the opposite side and then calculate their relevance scores using the backbone model. 
This estimation serves as an approximation of the $\bar{y}(a_i)$ and $\bar{y}(b_j)$ based on the sampled users. By considering both relevance and direction, the reranking process ensures rankings with the most matching expectations, taking into consideration the reciprocal nature of the RRS.

\subsection{Optimization}

\begin{algorithm2e}[h]
  \DontPrintSemicolon\SetNoFillComment
  \small
  \SetKwInOut{Input}{input}
  \caption{The learning algorithm of \textbf{CRRS}.}
  \label{algo:crrs}
  \Input{User set $\mathcal{A}$ and User set $\mathcal{B}$, interaction set $\mathcal{D}_{11}$ with bilateral treatment, $\mathcal{D}_{01}$ with treatment for $A$ side and $\mathcal{D}_{10}$ with treatment for $B$ side.}
  \BlankLine
  Pre-train the backbone model with relevance $r_{ij}$.\;
  Initialize three models under different treatment assignments with the same pre-trained user embeddings $\bm{a}_i$ and $\bm{b}_j$. ~ \# (Sec.~\ref{sec:pretrain})\;
  
  \ForEach{mini-batch $\mathcal{M}$}{
    $\mathcal{L}_t \gets 0, t \in \{11, 10, 01\}$ \;
    Calculate $\mathcal{L}_{pretrain}$ with $\mathcal{M}$.\;
    \ForEach {user-user pair $(a_i, b_j)\in \mathcal{M}$}{
    % \IF{}
        Random get a negative sample $b_{j'}$ for $A$ side and a negative sample $a_{i'}$ for $B$ side.\; 
        \If {$(a_i, b_j) \in \mathcal{D}_{01}$}
        {
            $\mathcal{L}_{01} \gets \mathcal{L}_{01} - \sigma\big(\hat{y}_{01}(a_i,b_j) - \hat{y}_{01}(a_{i},b_{j'})\big) $\;
        }
        \If {$(a_i, b_j) \in \mathcal{D}_{10}$}
        {
            $\mathcal{L}_{10} \gets \mathcal{L}_{10} - \sigma\big(\hat{y}_{10}(a_i,b_j) - \hat{y}_{10}(a_{i'},b_{j})\big) $\;
        }
        \If {($a_i, b_j) \in \mathcal{D}_{11}$}
        {
            $\mathcal{L}_{11} \gets \mathcal{L}_{11} - \sigma\big(\hat{y}_{11}(a_i,b_j) - \hat{y}_{11}(a_i,b_{j'})\big) $\;
            $\mathcal{L}_{11} \gets \mathcal{L}_{11} - \sigma\big(\hat{y}_{11}(a_i,b_j) - \hat{y}_{11}(a_{i'},b_{j})\big) $\;
            $\mathcal{L}_{01} \gets \mathcal{L}_{01} - \sigma\big(\hat{y}_{01}(a_i,b_j) - \hat{y}_{01}(a_{i},b_{j'})\big) $\;
            $\mathcal{L}_{10} \gets \mathcal{L}_{10} - \sigma\big(\hat{y}_{10}(a_i,b_j) - \hat{y}_{10}(a_{i'},b_{j})\big) $\;
        }
    }
    $\mathcal{L}_t \gets \mathcal{L}_t + \mathcal{L}_{pretrain}, t \in \{11, 10, 01\}$ \;
    Backpropagate each $\mathcal{L}_t$ and update the parameters \;
  }
\end{algorithm2e}

Our proposed CRRS approach consists of two learning stages: pre-training and counterfactual learning. 
In the pre-training stage, we use real matching labels to train a backbone recommender model.  
In the counterfactual learning stage, we finetune the recommender model with direction labels to better predict outcomes, taking into account the causal effects of various recommendation strategies.
The full algorithm is described in Algorithm~\ref{algo:crrs}.

\subsubsection{First Stage: Pre-training.} 
\label{sec:pretrain}
We optimize the backbone model using the Bayesian Personalized Ranking~(BPR) loss \cite{rendle2009bprmf}, which effectively captures the relative preferences between positive and negative samples. 
To describe the pre-training process, we 
take side $A$ for illustration when it is symmetric for side $B$, we define the pre-training loss $\mathcal{L}_{pretrain}$ as follows:

\begin{align}
\label{eq:pretraining}
    \mathcal{L}_{pretrain} = -\frac{1}{|{\mathcal{D}}|}\sum_{(a_i, b_j) \in {\mathcal{D}}}\log \big(\sigma\big(\hat{y}(a_i,b_j) - \hat{y}(a_i,b_{j'})\big) \big),
\end{align}
where $\sigma(\cdot)$ is the sigmoid function, $\mathcal{D}$ denotes the training data,  $(a_i, b_j)$ is a positive sample for matching, and $b_{j'}$ denotes the negative sample for user $a_i$. 
This stage aims to learn a basic recommender model for subsequent causal finetuning. 

\subsubsection{Second Stage: Counterfactual Learning.} 
At this stage, 
we further finetune the recommendation model using data under different treatment assignments. Similar to the first stage, the counterfactual learning loss also employs the BPR ranking loss. For each backbone model, the loss function is expressed as follows:

\begin{align}
\label{eq:finetuning}
    \mathcal{L}_{t} = -\frac{1}{|{\mathcal{D}_t}|}\sum_{(a_i, b_j) \in {\mathcal{D}_t}}\log \big(\sigma\big(\hat{y}_t(a_i,b_j) - \hat{y}_t(a_i,b_{j'})\big) \big),
\end{align}
where $\mathcal{D}_t$ represents the training data specific to each treatment. To ensure training stability, we combine the counterfactual learning loss with the pre-training loss. 
The treatment-specific data provides important guidance that instructs the models to refine their performance under specific conditions.  
In this way, the models can more effectively discern the causal effects between treatments and outcomes, thus better achieving better performance for RRS.

\subsection{Discussion}
We conduct a discussion here to clarify our stance on redundant recommendations.
In practice, different scenarios may exhibit varying stances toward redundant recommendations. 
For example, 
on LinkedIn\footnote{https://linkedin.com/}, mutual recommendations between matched candidates and recruiters are optional. In contrast, on Tinder\footnote{https://tinder.com/}, two users can only start communicating if they both recommend each other and click the ``like'' button.
In this paper, we adopt a negative stance towards redundant recommendations, prioritizing the enhancement of overall coverage and achieving higher quantities as fundamental objectives (measured by CRecall and CPrecision), despite potential trade-offs in stability (SRecall and SPrecision), as evidenced in subsequent experiments.

%% file: sec_experiment.tex
\section{Experiments}
\label{sec:experiment}
In this section, we aim to evaluate the proposed CRRS's performance and the proposed metrics' effectiveness.
We start with the experiment setup, followed by a thorough evaluation comparing the CRRS with baselines. Then we analyze traditional versus proposed metrics and present results from ablation studies alongside a ranking analysis of redundant recommendations.

\subsection{Experimental Setup}

\begin{table}[t]
\caption{Statistics of the experimental datasets.}
\label{tab:statistics_datasets}
\resizebox{0.48\textwidth}{!}{
\begin{tabular}{crrrrc}
\toprule
Dataset   & \#User A & \#User B & \#Interaction  & \#Match & Sparsity \\
\midrule
Recruitment      & $32,161$  & $25,665$  & $790,725$    & $224,636$  & $99.90\%$  \\
Dating & $6,391 $  & $6,516 $  & $605,288$ & $51,474 $  & $98.55\%$  \\
\bottomrule
\end{tabular}
}
\end{table}

\paratitle{Dataset.} To evaluate the effectiveness of our method, we conduct experiments on two large real-world datasets from different reciprocal recommendation domains: recruitment and online dating.
% \begin{itemize} [leftmargin=0.4cm]

% \item
$\bullet$ 
\textbf{Recruitment}~\cite{yang2022modeling}: This dataset contains two weeks of real logs collected from a large and popular online recruitment platform, which records dynamic interactions between job-seeking candidates and proactive recruiters.

% \item
$\bullet$ 
\textbf{Dating\footnote{\url{http://konect.cc/networks/libimseti/}}}: 
This dataset is collected from the popular online dating service Libimseti. 
It encompasses ratings between users, and we only keep ratings between opposite-sex pairs (\ie female-to-male and male-to-female). Pairs with mutual ratings of 8 or above are viewed as a successful match, given the increased likelihood of a match with bilateral exposure.

Following the standard practice~\cite{rendle2010factorizing, zhou2020s3rec},  we only keep the 5-core interactions for both datasets. After preprocessing, we randomly split both datasets into training, validation, and test sets with ratios of 8 : 1 : 1. The statistics of both datasets are summarized in~\autoref{tab:statistics_datasets}.

\begin{table*}
\centering
\caption{Performance comparison of all methods on Recruitment and Dating dataset. The best performance and the second best performance methods are denoted in bold and underlined fonts, respectively. ``${*}$'' indicates that the improvements are statistically significant with paired $t$-test ($p$-value < 0.05). \textcolor{black}{``True Positive Pairs'' represent the count of matching pairs.}
% \textcolor{red}{add a TP introduction}
}
\label{tab:overall_comparison}
\resizebox{0.98 \textwidth}{!}{
\begin{tabular}{ccl|cccc|cccc}
\toprule
\multirow{2}{*}{Dataset}& \multirow{2}{*}{Direction} & \multirow{2}{*}{Metric} & \multicolumn{4}{c|}{MF-based Model} & \multicolumn{4}{c}{Graph-based Model}  \\
& & & BPRMF& D-BPRMF & LFRR& CRRS (Ours) & LightGCN& D-LightGCN & DPGNN  & CRRS (Ours)  \\
\midrule
\multirow{15}{*}{Recruitment} & \multirow{3}{*}{$A$ side} & Recall@50 & \underline{$0.2424$} & $0.1223 $ & $\textbf{0.2569*} $ & $0.2309 $& $0.2688 $ & $0.1929 $ & $\textbf{0.3621*} $ & \underline{$0.3560 $}\\
& & Precision@50 & \underline{$0.0067 $} & $0.0035 $ & $\textbf{0.0071} $ & $0.0059 $& $0.0072 $ & $0.0053 $ & \underline{${0.0090} $} & $\textbf{0.0093} $\\
& & NDCG@50& $0.1070 $ & $0.0632 $ & $\textbf{0.1141*} $ & \underline{$0.1077 $}& $0.0997 $ & $0.0741 $ & $\textbf{0.1573*} $ & \underline{$0.1276 $}\\
 % & & MRR@50 & $0.0814 $ & $0.0556 $ & $\textbf{0.0874*} $ & \underline{$0.0842 $}& $0.0632 $ & $0.0486 $ & $\textbf{0.1130*} $ & \underline{$0.0768 $}\\
\cmidrule{2-11}
& \multirow{3}{*}{$B$ side} & Recall@50 & $0.1886 $ & $0.1282 $ & \underline{$0.2001 $} & $\textbf{0.2355*} $& $0.1764 $ & $0.1384 $ & $\textbf{0.3038*} $ & \underline{$0.2757 $}\\
& & Precision@50 & $0.0063 $ & $0.0043 $ & \underline{$0.0067 $} & $\textbf{0.0072*} $& $0.0056 $ & $0.0045 $ & $\textbf{0.0093*} $ & \underline{$0.0088 $}\\
& & NDCG@50& $0.0887 $ & $0.0606 $ & \underline{$0.0945 $} & $\textbf{0.1016*} $& $0.0625 $ & $0.0533 $ & $\textbf{0.1288*} $ & \underline{$0.0973 $}\\
% & & MRR@50 & $0.0762 $ & $0.0532 $ & $\textbf{0.0817} $ & \underline{$0.0803 $}& $0.0394 $ & $0.0373 $ & $\textbf{0.0972*} $ & \underline{$0.0607 $}\\
\cmidrule{2-11}
& \multirow{3}{*}{Average}& Recall@50 & $0.2155 $ & $0.1253 $ & \underline{$0.2285 $} & $\textbf{0.2332*} $& $0.2226 $ & $0.1656 $ & $\textbf{0.3329*} $ & \underline{$0.3158 $}\\
& & Precision@50 & $0.0065 $ & $0.0039 $ & $\textbf{0.0069*} $ & \underline{$0.0066 $}& $0.0064 $ & $0.0049 $ & $\textbf{0.0091} $ & \underline{$0.0090 $}\\
& & NDCG@50& $0.0978 $ & $0.0619 $ & \underline{$0.1043 $} & $\textbf{0.1046} $& $0.0811 $ & $0.0637 $ & $\textbf{0.1431*} $ & \underline{$0.1124 $}\\
% & & MRR@50 & $0.0788 $ & $0.0544 $ & $\textbf{0.0846*} $ & \underline{$0.0822 $}& $0.0513 $ & $0.0430 $ & $\textbf{0.1051*} $ & \underline{$0.0687 $}\\
\cmidrule{2-11}
& \multirow{6}{*}{Overall}& SRecall@50 & \underline{$0.0617 $} & $0.0152 $ & $\textbf{0.0967*} $ & $0.0242 $& $0.0638 $ & $0.0414 $ & $\textbf{0.1535*} $ & \underline{$0.1248 $}\\
& & SPrecision@50 & \underline{$0.0021 $} & $0.0005 $ & $\textbf{0.0029*} $ & $0.0006 $& $0.0018 $ & $0.0013 $ & $\textbf{0.0042*} $ & \underline{$0.0036 $}\\
& & RNDCG@50  & $0.0989 $ & $0.0620 $ & $\textbf{0.1054} $ & \underline{$0.1050 $}& $0.0832 $ & $0.0649 $ & $\textbf{0.1447*} $ & \underline{$0.1142 $}\\ \cmidrule{3-11}
& & \textbf{CRecall@50} & $0.3388 $ & $0.2336 $ & \underline{$0.3530 $} & $\textbf{0.3968*} $& $0.3609 $ & $0.2811 $ & \underline{$0.4555 $} & $\textbf{0.4670*} $\\
& & \textbf{CPrecision@50} & $0.0052 $ & $0.0036 $ & \underline{$0.0054 $} & $\textbf{0.0061*} $& $0.0056 $ & $0.0043 $ & \underline{$0.0070 $} & $\textbf{0.0073*} $\\

& & \textbf{True Positive Pairs}  & 7,610  & 5,247  & \underline{7,929}  & \textbf{8,913*}& 8,106  & 6,314  & \underline{10,231}  & \textbf{10,490*}\\
\midrule
% \midrule
\multirow{15}{*}{Dating}& \multirow{3}{*}{$A$ side} & Recall@50 & \underline{0.1760} & 0.1211 & 0.1738 &                        \textbf{0.1783*}    & 0.1787 & 0.1719 & \underline{0.1888} & \textbf{0.2172*} \\
& & Precision@50 & 0.0073 & 0.0053 & \underline{0.0074} &          \textbf{0.0077*}                   & 0.0070 & 0.0071 & \underline{0.0076} & \textbf{0.0088*} \\
& & NDCG@50& \underline{0.0756} & 0.0052 & 0.0719 &      \textbf{0.0809*}       & 0.0769 & 0.0724 & \underline{0.0788} & \textbf{0.0922*} \\
\cmidrule{2-11}
& \multirow{3}{*}{$B$ side} & Recall@50 & 0.1343 & 0.0830 & \textbf{0.1719} &    \underline{0.1607}      & 0.1233 & 0.1616 & \underline{0.1762} & \textbf{0.1916*} \\
& & Precision@50 & 0.0069 & 0.0047 & \textbf{0.0083} &      \underline{0.0082}     & 0.0059 & 0.0077 & \underline{0.0079} & \textbf{0.0089*} \\
& & NDCG@50& 0.0566 & 0.0360 & \textbf{0.0693} &         \underline{0.0692}          & 0.0499 & 0.0676 & \underline{0.0689} & \textbf{0.0777*} \\
\cmidrule{2-11}
& \multirow{3}{*}{Average}& Recall@50 & 0.1552 & 0.1021 & \textbf{0.1729} &    \underline{0.1695}     & 0.1510 & 0.1668 & \underline{0.1825} & \textbf{0.2044*} \\
& & Precision@50 & 0.0071 & 0.0050 & \underline{0.0079} &     \textbf{0.0080}   & 0.0065 & 0.0074 & \underline{0.0078} & \textbf{0.0089*} \\
& & NDCG@50& 0.0661 & 0.0206 & \underline{0.0706} &    \textbf{0.0751}    & 0.0634 & 0.0700 & \underline{0.0739} & \textbf{0.0850*} \\
\cmidrule{2-11}
& \multirow{6}{*}{Overall}& SRecall@50 & 0.0830 & 0.0419 & \underline{0.0968} &        \textbf{0.0993*}          & 0.0753 & 0.0732 & \underline{0.0997} & \textbf{0.1221*} \\
& & SPrecision@50 & 0.0018 & 0.0009 & \underline{0.0021} &       \textbf{0.0022}        & 0.0017 & 0.0016 & \underline{0.0022} & \textbf{0.0027*} \\
& & RNDCG@50  & 0.0660 & 0.0207 & \underline{0.0706} &   \textbf{0.0751*}      & 0.0633 & 0.0700 & \underline{0.0738} & \textbf{0.0849*} \\  \cmidrule{3-11}
& & \textbf{CRecall@50} & 0.2795 & 0.2055 & \underline{0.3045} & \textbf{0.3086*}   & 0.2549 & 0.2986 & \underline{0.3007} & \textbf{0.3387*} \\
& & \textbf{CPrecision@50} & 0.0062 & 0.0045 & \underline{0.0067} &       \textbf{0.0068}          & 0.0056 & 0.0066 & \underline{0.0067} & \textbf{0.0075*} \\

& & \textbf{True Positive Pairs}  &  1,439  &  1,058  & \underline{1,567 } & \textbf{1,588*}& 1,312  & 1,537  & \underline{1,548 } & \textbf{1,743*}\\
\bottomrule
\end{tabular}}
\end{table*}

\paratitle{Baseline models.} As our proposed CRRS is model-agnostic, we can apply it to various recommendation models. We choose the following two popular models as our backbone models:
 
$\bullet$ 
{\textbf{BPRMF}~\cite{rendle2009bprmf}}: This model designs Bayesian Personalized Ranking~(BPR) loss to optimize the Matrix Factorization (MF), which is representative in the collaborative filtering recommendation.
 
$\bullet$ 
{\textbf{LightGCN}~\cite{he2020lightgcn}}: This model designs a simple but competitive GCN architecture for the graph-based recommendation.

We also compare our proposed CRRS with
the following reciprocal recommendation models:

$\bullet$ 
\textbf{D-BPRMF}: This method leverages two same BPRMF but with different parameters to make recommendations for each side.

$\bullet$ 
{\textbf{LFRR}~\cite{neve2019latent}}: LFRR is a latent factor model designed specifically for reciprocal recommendation scenarios, which aggregates two MFs to model the preferences of both sides involved.

$\bullet$ 
\textbf{D-LightGCN}: This method is similar to D-BPRMF, but it utilizes LightGCN as the backbone model.

$\bullet$ 
{\textbf{DPGNN}~\cite{yang2022modeling}}: DPGNN is a graph-based method to model the two-way selection preference through a dual-perspective graph representation learning approach.

Note that these baseline models fall into two categories: MF-based Models and graph-based Models, where the backbone models used are BPRMF and LightGCN, respectively. 
The models with the ``D-'' prefix employ two identical backbone models to train and make predictions independently for each side. LFRR and DPGNN can be viewed as extensions of BPRMF and LightGCN to reciprocal recommendation scenarios, respectively.
To be more inclusive, we just opted for models in the general setting rather than those designed specifically for sequences~\cite{zheng2023reciprocal, chen2023bilateral}, which have more complex data and model requirements.
% However, our approach can also be adapted to this scenario.}

\paratitle{Evaluation settings.} 
To evaluate the performance of the top-$K$ recommendation of RRS, we employ both traditional metrics Recall, Precision, NDCG, and metrics we proposed in Section~\ref{sec:metric}.
For each dataset, we employ a full-ranking evaluation where every candidate user is ranked based on its relevance for the target user.
We first report on single-sided metrics, treating RRS as two independent ranking tasks: ranking all users from side $B$ for each user $a_i \in \mathcal{A}$ and vice versa.
Meanwhile, we record the average values of the single-sided metrics.
Following this, we record the five proposed metrics: CRecall, CPrecision, SRecall, SPrecision, and RNDCG, to measure the comprehensive capability of the models.
Additionally, we report the number of matching pairs
under recommendations~(denoted as True Positive Pairs).

\paratitle{Implementation details.}
We implement baseline models using a popular open-source recommendation library \textsc{RecBole}~\cite{zhao2021recbole, zhao2022recbole}.
For a fair comparison, the dimension of user embedding is standardized to 128 across all models.
$K$ is set to 50 empirically.
Moreover, we optimize all methods using the Adam optimizer and conduct a thorough search for the hyper-parameters of all baselines. 
The learning rate is tuned from \{0.001, 0.0001, 0.00001\} for optimal performance. 
To prevent overfitting, we employ early stopping with patience of 30 epochs.
To facilitate evaluations in offline experiments, we consider that all positive instances can be successfully matched when recommendations are made exclusively on one side during both the training and evaluation phases.

\subsection{Experimental Results}

\subsubsection{Overall Performance.}
We conduct a comprehensive comparison of different methods across various evaluation metrics.
The results on two datasets are reported in Table~\ref{tab:overall_comparison}. 

\paratitle{Performance comparison between models.}
As we can see, in \textit{Recruitment}, reciprocal methods LFRR and DPGNN demonstrate superior performance across most single-sided evaluation metrics, due to their ability to effectively capture the underlying interactions. 
Our proposed CRRS is not necessarily better than a well-designed reciprocal recommendation model in single-sided metrics but achieves the highest performance in overall coverage metrics $\text{CRecall}$ and $\text{CPrecision}$ and the count of matching pairs. 
For the \textit{dating}, owing to the behavioral direction characteristics in the data and there being relatively more user interactions, the proposed method outperforms others across all metrics. This can be attributed to the continued training stage, which enables the model to better capture the correlations between users on both sides.
Moreover, the D-BPRMF and D-LightGCN exhibit almost the worst performance across all metrics, largely due to their insufficient training data.

\paratitle{Trade-off between coverage and stability.}
In \textit{Recruitment} dataset, a phenomenon emerges when comparing CRRS with the best baseline model DPGNN: there is a rise in CRecall~(0.4670 \emph{v.s.} 0.4555), contrasted by a fall in SRecall~(0.1248 \emph{v.s.} 0.1535), similar to CPrecision and SPrecision. This phenomenon is normal when trying to reduce redundant recommendations as analyzed in Section~\ref{sec:metric_analysis}.
However, this phenomenon does not occur in \textit{dating} dataset, possibly because there is a significant difference in sparsity between the two datasets on ``\#Match'' (99.90\% \emph{v.s.} 98.55\%), with even greater sparsity differences on ``\#Interactions''.
The inherent conflict between coverage and stability exists due to the different stances towards redundant recommendations, suggesting that trade-offs may be required in certain situations.

\subsubsection{Metric Comparative Analysis}

\begin{figure*}[t]
	\centering
    \includegraphics[width=0.97\textwidth]{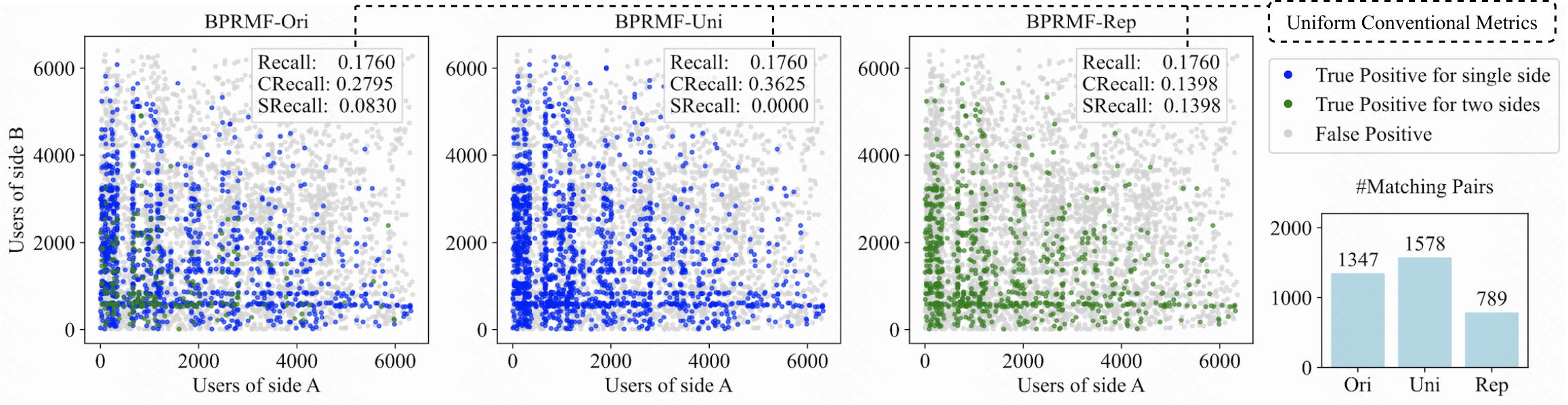}
	\caption{\textcolor{black}{Comparative analysis between traditional and proposed metrics.
 We employ the same model BPRMF on \textit{Dating} and adjust the redundant recommendations by manually converting them into non-redundant ones and vice versa. 
 In three cases, the models demonstrate identical traditional metrics yet display completely divergent overall performance.}}
	\label{fig:manual_exp}
\end{figure*}

We conduct experiments to compare traditional metrics and the proposed metrics under different recommendation strategies on \emph{dating}.
For the same model, we manually adjust the redundant recommendations and consider the following two adjustments: \textbf{(A)} \underline{BPRMF-Uni}: Convert the redundant recommendations with another matching pair that does not appear on both sides. \textbf{(B)} \underline{BPRMF-Rep}: Convert the non-redundant pairs with other matching pairs that have appeared on one side. The original model is named as \underline{BPRMF-Ori}. As shown in Figure~\ref{fig:manual_exp}, We depict a rectangular interaction diagram with users from both sides represented along the horizontal and vertical axes.
Blue points are true positives for just one side and the green ones are true positives for the two sides.
Recommendations on three cases have completely identical traditional ranking metrics~(Recall, Precision, and NDCG). 
However, significant differences exist in the proposed metrics, which directly influence the count of matching pairs. 
Thus, it is critical to take the proposed metrics into account in RRS.

\subsubsection{Ablation Study}

\begin{table}[]
\label{tab:ablation_study}
\caption{Effectiveness analysis of key components of CRRS.
CR, CP, SR, and SP are the simplifications for CRecall, CPrecision, SRecall, and SPrecision, respectively.
}
\label{tab:ablation_study}
\resizebox{0.48\textwidth}{!}
{

\begin{tabular}{ccccccc}
\toprule
\multirow{2}{*}{Dataset}     & \multirow{2}{*}{Model} & \multicolumn{5}{c}{Proposed Metric}                                                     \\ \cmidrule(l){3-7} 
                             &                        & CR@50           & CP@50           & SR@50           & SP@50           & RNDCG           \\ \midrule
\multirow{5}{*}{Recruitment} & CRRS~(BPRMF)      & \textbf{0.3968} & \textbf{0.0061} & 0.0242          & 0.0006          & 0.1050          \\
                             & w/o finetune           & 0.3694          & 0.0057          & 0.0397          & \textbf{0.0023} & \textbf{0.1070} \\
                             & w/o pre-train          & 0.3400          & 0.0052          & 0.0527          & 0.0014          & 0.0689          \\
                             & w/o rerank             & 0.3091          & 0.0048          & \textbf{0.0757} & 0.0020          & 0.0882          \\
                             & BPRMF                  & 0.3388          & 0.0052          & 0.0617          & 0.0021          & 0.0989          \\ \midrule
\multirow{5}{*}{Dating}      & CRRS~(BPRMF)      & \textbf{0.3086} & \textbf{0.0068} & 0.0993          & 0.0022          & \textbf{0.0751} \\
                             & w/o finetune           & 0.2927          & 0.0065          & 0.0696          & {0.0015} & {0.0692} \\
                             & w/o pre-train          & 0.2986          & 0.0066          & 0.0826          & {0.0018} & 0.0669          \\
                             & w/o rerank             & 0.3011          & 0.0067          & \textbf{0.1123} & \textbf{0.0025} & 0.0748          \\
                             & BPRMF                  & 0.2795          & 0.0062          & 0.0830           & 0.0018          & 0.066           \\ \bottomrule
\end{tabular}
}
\end{table}

The major technical contribution of our approach lies in the potential outcome framework training algorithm, as well as the proposed reranking strategy. 
We now analyze how each part contributes to the final performance.
We consider the following three variants of CRRS: \textbf{(A)} \underline{CRRS w/o finetuning}~(Equation~\eqref{eq:finetuning}); \textbf{(B)} \underline{CRRS w/o pre-training}~(Equation~\eqref{eq:pretraining}); \textbf{(C)} \underline{CRRS w/o} \underline{reranking}~(Equation~\eqref{eq:reranking});

% \textcolor{blue}{
In Table~\ref{tab:ablation_study}, we can observe that the absence of any component will lead to performance degradation, with varying impacts across different datasets.
In \textit{recruitment}, the reranking strategy contributes significantly to the overall coverage improvement as its objective is to avoid redundant recommendations. 
At the same time, we can observe that the result of the model without reranking demonstrates superior performance in bilateral stability.
In \textit{dating}, pre-training and finetuning play more vital roles in enhancing the model performance across various metrics. 
While the benefits brought by the reranking strategy are relatively modest.
These results highlight the usefulness of all components in CRRS.
% }

% \subsection{Further Analysis}

\subsubsection{Ranking Analysis of Redundant Recommendations}
\begin{figure}[t!]
	\centering
    \includegraphics[width=0.47\textwidth]{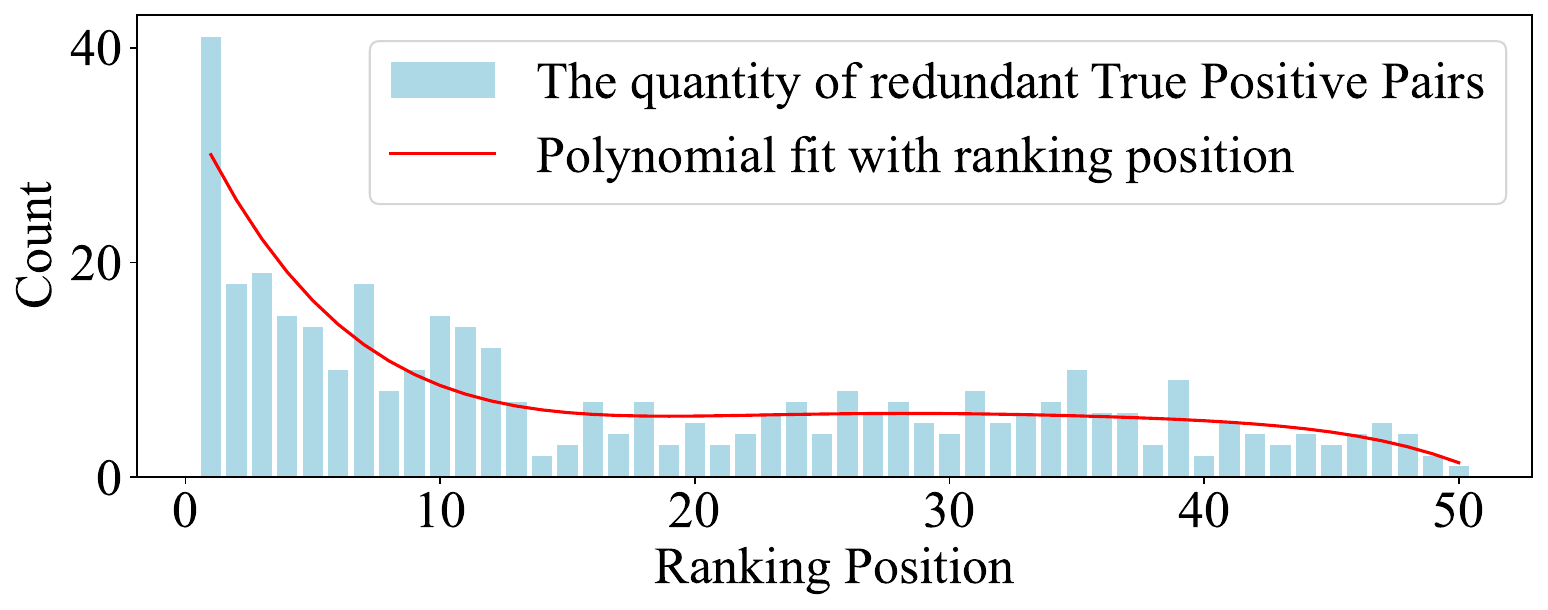}
	\caption{\textcolor{black}{Distribution of ranking positions for redundant recommendations within the top 50 positions. 
 }}
	\label{fig:ranking}
\end{figure}

We conduct experiments here to analyze the distribution of redundant recommendations in the ranking list.
We employ the same BPRMF model to generate top-$50$ recommendations for both sides and record the ranking of all redundant recommendations.
\textcolor{black}{As discussed above}, an effective way to enhance coverage involves reducing duplication between the bilateral ranking lists. 
However, as Figure~\ref{fig:ranking} shows, the majority of redundant recommendations are found in the advanced positions, so an objective of directly optimizing the discrepancy between both sides may be ineffective.
This suggests that innovative approaches like reranking are essential for achieving more effective and balanced outcomes.
% (5th to 10th is easy, but 5th to 60th may be arduous). 
% Consequently, we design a reranking strategy based on predictions of potential outcomes, rather than directly optimizing both tasks as a unified objective.

% \subsubsection{Repeat Pair Analysis}

%% file: sec_conclusion.tex
\section{Conclusion}
\label{sec:conclusion}
This paper revisited the study of reciprocal recommender systems by introducing new metrics, formulation, and method. 
Firstly, we proposed  
five metrics for evaluating RRS from three new aspects.
% including overall coverage, bilateral stability, and balanced ranking.
Then we formulated reciprocal recommendation tasks from a causal perspective, considering the recommendations as bilateral interventions.
Furthermore, we proposed a causal reciprocal recommendation model using a potential outcome framework.
% to capture the causal effects of recommendations.
To optimize overall performance, we further designed a reranking strategy to enhance the overall performance.
Extensive experiments on two datasets indicate that the proposed approach can achieve superior performance on overall performance.
% compared with competitive baselines.